\documentclass[reprint, aps, pra]{revtex4-1}
\usepackage{graphicx}
\usepackage{amsfonts}
\usepackage{amssymb}
\usepackage{amsthm}
\usepackage{array}
\usepackage{amsmath}
\usepackage{verbatim} 
\usepackage{hyperref}
\usepackage{color}
\usepackage{bbold}
\usepackage{epstopdf}
\usepackage{mathtools}

\newcommand{\ket}[1]{\left\vert#1\right\rangle}
\newcommand{\bra}[1]{\left\langle#1\right\vert}

\def\bra#1{\langle #1|}
\def\ket#1{\left|#1 \right>}

\begin{document}
\title{Comment on ``Quantifying quantum coherence with quantum Fisher information"}
\author{Hyukjoon Kwon}
\affiliation{Center for Macroscopic Quantum Control, Department of Physics and Astronomy, Seoul National University, Seoul, 151-742, Korea}
\author{Kok Chuan Tan}
\affiliation{Center for Macroscopic Quantum Control, Department of Physics and Astronomy, Seoul National University, Seoul, 151-742, Korea}
\author{Seongjeon Choi}
\affiliation{Center for Macroscopic Quantum Control, Department of Physics and Astronomy, Seoul National University, Seoul, 151-742, Korea}\author{Hyunseok Jeong}
\affiliation{Center for Macroscopic Quantum Control, Department of Physics and Astronomy, Seoul National University, Seoul, 151-742, Korea}
\date{\today}
\begin{abstract}
We show that  contrary to the claim in Sci. Rep. {\bf 7}, 15492 (2017),
the quantum Fisher information itself is not a valid coherence measure based on the resource theory of coherence because it can increase via an incoherent operation.
\end{abstract}
\pacs{}
\maketitle
In Ref.~\cite{Feng17}, the authors claim that the quantum Fisher information (QFI) is a coherence measure in the sense of  the resource theory of coherence suggested in Ref.~\cite{Baumgratz14}. This means that the  QFI is a monotone under an incoherent operation.  
%We adopt the same notations introduced in Ref.~\cite{Feng17} and review the (flawed) statement.
%{\bf What have they done?}

The resource theory of coherence \cite{Baumgratz14} with respect to a fixed set of basis $\{ \ket{i} \}$ can be constructed by 
a set of incoherent states $\delta \in \Pi$ that contain only diagonal components, i.e., $\delta = \sum_i p_i \ket{i}\bra{i}$
and a set of incoherent operations $\Phi$ which map every incoherent state into another incoherent state, i.e., $\Phi(\Pi) \subseteq \Pi$.
Then a coherence measure $C(\rho)$ for state $\rho$ should satisfy the following conditions \cite{Baumgratz14}:
(C1) $C(\rho) \leq 0$ and $C(\rho) = 0$ iff $\rho \in \Pi$.
(C2a) Non-increasing under an incoherent completely positive and trace preserving operation $\cal O_ I$, i.e., $C(\rho) \geq C (\cal O_ I[\rho])$.
(C2b) Non-increasing on average by selective incoherent operations, i.e., $C(\rho)\geq \sum_n p_n C(A_n \rho A_n^\dagger / p_n)$, where $\Phi(\rho) = \sum_n A_n \rho A_n^\dagger$ and $p_n = {\rm Tr} \rho A_n^\dagger A_n$.
(C3) Convexity $\sum_n p_n C(\rho_n) \geq C(\sum_n p_n \rho_n)$.

The authors of Ref.~\cite{Feng17} claim that the QFI with respect to a given Hamiltonain $H$
$$F (\rho, H) = 2 \sum_{i,j} \frac{(\lambda_i - \lambda_j)^2}{\lambda_i + \lambda_j} |\langle \lambda_i | H |\lambda_j \rangle|^2$$
satisfies all the conditions for coherence measure (C1)--(C3) with respect to the eigenbasis of the Hamiltonian $H$, where $\lambda_i$ and $\ket{\lambda_i}$ are eigenstates and eigenvalues of the quantum state $\rho$, respectively.

However, the proof of (C2) is incorrect. There exists a counterexample in which an incoherent operation can increase the QFI.
We consider a Hamiltonian in an $N$-level system with equal energy spacing,
$$H = \sum_{n=1}^N n \ket{n} \bra{n}.$$
Assume that a  quantum state $\ket{\psi}$ is initially given by 
\begin{equation}
\ket{\psi} = \frac{1}{\sqrt{2}} \left( \ket{0} + \ket{1}\right).
\end{equation}
We consider an incoherent unitary operation
$$
U = \ket{N}\bra{1} + \ket{1}\bra{N}+ \ket{0}\bra{0} + \sum_{n=2}^{N-1} \ket{n}\bra{n}
$$
which simply exchanges $\ket{1}$ and $\ket{N}$, while leaving the other eigenstates  unchanged.
It is important to note that $U$ as an incoherent operation maps any incoherent state into another incoherent state.

Under this incoherent unitary $U$, $\ket{\psi}$ evolves into
$$
U \ket{\psi} = \frac{1}{\sqrt{2}} \left( \ket{0} + \ket{N}\right).
$$
Using the fact that the QFI of a pure state equals to four times of the variance of $H$, we can show that the QFI before and after the incoherent unitary $U$ is given by
$F (\ket{\psi}, H) = 1$ and $F (U \ket{\psi}, H) = N^2$, respectively.
It is thus immediately clear that the QFI can increase though an incoherent operation, and  $F(U \ket{\psi}, H) > F(\ket{\psi}, H)$ for every $N>1$.

%In addition, we point out that condition (C1)  may not be valid when $H$ is degenerate because of  the plurality of  basis choices. In this case, a quantum state can have a nonzero coherence between degenerate eigenstates while the QFI is zero. For example, suppose that the variable $g_0$ is an index corresponding to the degenerate states of some energy level $E_0$. We then find that the QFI of state $\ket{\phi} = \alpha \ket{E_0, g_0} + \beta \ket{E_0, g_0'}$ is zero while  the coherence is nonzero between the degenerate energy eigenstates when $H = \sum_{n}\sum_{g_n,g_n'} E_n \ket{E_n, g_n} \bra{E_n, g_n'}$. This result may change when a different basis is chosen.

\end{document}